\documentstyle[12pt,psfig]{article}

\def\ni{\noindent}
\def\be{\begin{equation}}
\def\ee{\end{equation}}

\begin{document}
\setlength{\textheight}{7.7truein}  %for 2nd page onwards

\thispagestyle{empty}

\setcounter{page}{1}

\vspace*{0.88truein}

%\fpage{1}
\centerline{\bf Roughness Scaling of Deconstruction}
\vspace*{0.035truein}
\centerline{\bf Interfaces.}
\vspace*{0.37truein}
\centerline{\footnotesize JUAN R. SANCHEZ
\footnote{Email: jsanchez@fi.mdp.edu.ar}}
\baselineskip=12pt
\centerline{\footnotesize\it Departamento de F\'{\i}sica,
Facultad de Ingenier\'{\i}a}
\baselineskip=10pt
\centerline{\footnotesize\it Universidad Nacional de Mar del plata}
\baselineskip=10pt
\centerline{\footnotesize\it Av. J.B. Justo 4302, 7600 Mar del Plata,
Argentina}
\vspace*{0.225truein}

\vspace*{0.25truein}
\abstract{The scaling properties of one-dimensional deconstructed 
surfaces are studied by numerical simulations of a disaggregation model.
The model presented here for the disaggregation process takes into
account the possibility of having quenched disorder in the bulk under
deconstruction.
The disorder can be considered to model several types
of irregularities appearing in real materials (dislocations, impurities).
The presence of irregularities makes the intensity of the attack 
to be not uniform.
In order to include this effect, the computational bulk is
considered to be composed by two types of particles. 
Those particles
which can be easily detached and other particles that are not sensible to 
the etching attack.
As the detachment of particles proceeds in time, 
the dynamical properties of the rough interface are studied.
The resulting one-dimensional surface show   
self-affine properties and the values of the scaling exponents 
are reported when the interface is still moving near the depinning
transition. 
According to the scaling exponents presented here,
the model must be considered to belong to a new universality class.}{}{}

\vspace*{5pt}

%% =====================================================================
\vspace*{1pt}
\section{Introduction}		
\vspace*{-0.5pt}
\noindent
There are three fundamental physical processes that gives rise to the
morphology of a surface: deposition, surface diffusion and desorption.
In the past decade, the characteristics of the interfaces 
generated by the combination of deposition and surface diffusion
has been well studied.~\cite{fv-b,v-b,bs-b}
In particular for growth models, particles are added to the
surface and then they are allowed to relax by different mechanisms.
Many of this models have been shown
to lead to the formation of self-affine surfaces, characterized by
scaling exponents.
From a theoretical point of view, the studies dedicated to the
self-affine interfaces generated by growth models
can be considered to follow two main branches.
The studies about the properties of discrete models and the studies
about continuous models.
Discrete models where dedicated mainly to the study of the properties of
computational models in which the growth proceeds on an initially empty
lattice representing a d-dimensional substrate.
At each time step, the height of the lattice sites is increased 
by units (usually one unit) representing the incoming particles.
Different models only differs on the relaxation mechanisms proposed 
to capture specific experimental characteristics.
Then, the models are classified according to the values of the scaling 
exponents in several universality classes.
The continuous models,~\cite{vill} on the other hand, are based on stochastic
differential equations of the type

\begin{equation}
\label{pde1}
\partial h/\partial t = F - \nabla .~{\mathbf{j}} + noise \:,
\end{equation}

\ni in which $h({\mathbf{r}} ,t)$ is the thickness of the film deposited onto the
surface during time $t$, $F$ denotes the deposition rate and
$\mathbf{j}$ denotes the current along the surface which in turn depends
on the local surface configuration. 
If the surface current $\mathbf{j}$ represents some experimental nonlinear 
equilibrium processes it may gives rise to different types of
nonlinear terms.
The {\it noise} term corresponds to the fluctuations in the growth rate 
and, in general, is assumed to be uncorrelated.
The differential equations are solved, either numerically or
analytically (when it
is possible), and the scaling exponents are determined.
If the values of the exponents
are similar to those of the discrete models, it is said that both 
(discrete and continuous models) belong to the same
universality class. Although, the formal connection between both
approaches (discrete and continuous) is still an open question.

Other type of models in which the interface shows dynamic scaling properties
are those related with fluid invasion in porous media. In this case,
the noise term in Eq.~\ref{pde1} takes the form 
$\eta({\mathbf {r}}, h({\mathbf{r}} ,t))$, which is
known as {\it quenched noise} and is considered to be uncorrelated. 
Usually, the corresponding discrete computational 
models resume all kind of irregularities that a appear in real porous media 
by invading a computational porous media which has only blocked 
and non-blocked sites. 
These models are classified according to its own universality
classes in the same way that others growth models.

Desorption or detaching processes, despite its technological importance,
have  not received the same amount of attention. 
Maybe because they were considered to be, simply, the
reciprocal of the growth models and no new characteristics were
expected to appear.

There are many technological processes depending on the details of the
etching phenomena and only recently few of them have been modeled in 
order to understand the phenomena at an atomic level.~\cite{will}
For instance, corrosion is among the most important disaggregation
processes because its economical importance.
In this sense a simple deconstruction model was just introduced by
Reverberi and Scalas (RS).\cite{scalas} They found that their model is in fact
the reciprocal of a growth model known as Wolf-Villain (WV) model,\cite{wolf} 
and that the interface show the same characteristic exponents.
In this paper we present a discrete disaggregation model, which can be
considered as a generalization of the above mentioned RS 
deconstruction model.
Besides its importance from an experimental point
of view, academically, it is worth to study a deconstruction model in which the
attack process may take place upon an heterogeneous material, because
its ubiquity in many technological and applied areas.

%% ========================================================================
\vspace*{1pt}
\section{Theory}		
\vspace*{-0.5pt}
\noindent
Following previous studies realized on growth models, it is expected that the
deconstructed surface will present self-affine properties.
According to the rules depicted below for the computational model, no overhangs
will be generated and the surface will be of the solid-on-solid (SOS) type.
Then, the characteristics of interest are the
scaling properties of the width (or roughness) defined as
\be
\omega(L,t) = \left[\langle h^2(x,t)\rangle_x - \langle
h(x,t)\rangle_x^2\right]^{1/2}\: ~,
\ee
\ni where $\langle \cdot \rangle_x$ denotes ensemble average over the entire
lattice size
$L$, $h(x,t)$ is the height (measured from the bottom) of the column 
at position $x$, $\langle h(x,t)\rangle_x$ is
the average surface height and $t$ is the time, roughly proportional
to the number of removed monolayers $m$.

As for a growth processes we assume that the roughness obeys
the Family-Vicsek scaling relation,~\cite{fv01} valid for 
SOS self-affine interfaces,

\be
\omega(L,t) = L^{\alpha}\:f\left(\frac{t}{L^{z}}\right) ~,
\ee

\ni being $\alpha$ the roughness exponent because  $\omega \sim
L^\alpha$ when $t \gg L^{z}$ and $\beta$ the growth exponent
because $\omega \sim t^{\beta}$ when $t \ll L^{z}$. 
Of course, in the case studied here there is
no {\it growth}, but an exponent $\beta$ can be defined 
in the same way. $z = \alpha/\beta\:$ is the dynamic exponent.
From a computational point of view, there are various ways in which
the scaling exponents can be calculated.
The exponent $\beta$ is usually determined by measuring the slope of the
log-log plot of $\omega(L,t)$ vs $t$ at early times.
For the exponent $\alpha$, several techniques are available.
In discrete models, simulations with different systems sizes $L$
can be made. Then the values of the saturation width, ($\omega_s=\omega(L,t)$
for $t \gg L^{z}$), are plotted against $L$ and the 
exponent $\alpha$ is determined directly from the plot.
On the other hand, the height-height correlation 
function defined as
$c(r,\tau) = \langle [h(x,t) - h(x+r,
t+\tau)]^2\rangle^{1/2}$, can be used.
The correlation function is known to scale in the same way 
as the width $w(L,t)$; in particular
for $t \gg L^{z}$ and for $r \ll L$ the correlation function behaves as
$c(r,0) \sim r^\alpha \:$.
A third technique for calculating the dynamic exponent
$\alpha$ is based in the scaling properties of the
structure factor.~\cite{bs-b} Within this approach the dynamic scaling 
behavior can be investigated by
calculating the structure factor (power spectrum) defined as
\begin{equation}
S(k,t)=
\langle {h}(k,t) {h}(-k,t) \rangle \quad ,
\end{equation}
where ${h}(k,t)=({L^{-1/2}})\sum_x[h(x,t)-\overline
{h(t)}]e^{ikx}$.
The structure factor scales as
\begin{equation}
\label{struct1}
S(k,t)=k^{-(2\alpha+1)} g(t/k^{-z}) ~~,
\end{equation}
where the scaling function has the form
\begin{equation}
\label{struct2}
g(u)=
\left\{ \begin{array}{lcl}
u^{(2\alpha + 1)/z}  & u \ll 1 ~~, \\
const & u \gg 1 ~~.
\end{array}
\right.
\end{equation}
The equation $\omega^2(L,t) = (1/L) \sum_k S(k,t)$
gives the relation between the roughness and the structure factor.
\footnote{Recently,~\cite{rama} a generalization of the scaling 
properties of S(k,t) has been proposed. However here it is considered
that the Family-Vicsek scaling anzat is obeyed and hence the scaling
properties of the structure factor must be those represented by
equations~(\ref{struct1}-\ref{struct2}).}

It is worth to mention that for $t \gg \L^z$, the structure
factor does not show any dependence with the lattice size $L$.
This characteristic makes $S(k)$ very suitable to calculate the
exponent $\alpha$ using smaller system sizes for the simulations.
This approach will be used here.

%% =====================================================================
\vspace*{1pt}
\section{Computational Model}
\vspace*{-0.5pt}
\noindent
As stated above, the deconstruction model presented by Reverberi and Scalas
can be considered as the reciprocal of the WV growth model and shows the
same values of the scaling exponents. 
In the RS model the deconstruction proceeds by
picking at random a site in the lattice and remove the particle which
presents the smaller amount of bonds between the selected one and its
nearest neighbors. On the other hand, in the WV model the deposition 
process is as follows; after selecting at random a site, the number of 
bonds of the site is calculated as well as the number of bonds 
of it nearest neighbors sites.
The particle is deposited in the site with the larger amount of bonds.
This deposition mechanism is justified by saying that the number of bonds of
a site is {\it roughly} proportional to the {\it local chemical
potential}, because the number of bonds increases with the local curvature
of the interface, i.e., $\mu(x,t) \propto -\nabla^2 h(x,t)$.
In this sense, another growth model can be formulated. This model is
called the {\it larger curvature} (LC) model and was introduced by
Kim and Das Sarma.~\cite{kds} In this case, the larger local curvature
$[h(x+1)+h(x-1)-2 h(x)]$ between a random selected site and its 
neighbors is used to decide where
to deposit the new incoming particle. It can be shown, by using an argument
based on a Hamiltonian, that the LC  model belongs to the same universality
class that the WV model.~\cite{kds}

Following this line of reasoning, a deconstruction model equivalent to the RV model
would be one in which particles are removed from the site with
the {\em smaller} value of the local curvature. This would be the reciprocal
of the LC growth model. Lets call it the {\em smaller curvature} (SC)
deconstruction model.
Of course, by the same arguments stated above, the RV and the SC 
deconstruction models will belong to the same universality class; the
WV and LC universality class.

The model presented here is mostly related with another type of growth
models.
Those in which the advance of the interface takes place on
a random environment.\footnote{However, as it is shown below, the model 
presented here is not the direct reciprocal of any of the known 
growth models.} 
Several models of this kind has been
presented and their universality classes and scaling
exponents has been studied.~\cite{v-b,bs-b}
In those models, the growing interface advances in a random
environment. Usually, the randomness of a real porous media is
resumed in a computational random media in which there are
a certain amount of sites which are blocked to the advance 
of the interface while others are not.~\cite{bs-b,bul} 
Beyond a certain amount of blocked sites
present in the system the interface becomes {\it pinned} and
cannot advance any more. The scaling properties of such
interfaces is, in general, studied near the pinning transition.

Here, and in order to model various types of real heterogeneities, 
the deconstruction is considered to take place on a bulk 
in which two types of particles are present, those particles 
susceptible to the etching and those which cannot be 
detached by the external agent acting upon the interface. 
Then, the deconstruction rules run as follows. 
A surface is represented by a one-dimensional array of 
integers specifying the number of atoms in each one of the columns of the bulk.
To maintain the solid-on-solid characteristic of the model, the
detaching processes will not give rise to overhangs.
Each column of the one-dimensional array (indexed by $i$) is initially 
filled enough in order to allow the deconstruction process to take 
place during the whole simulational time, i.e., no column height must
become negative at the end of the simulation.
Before starting the deconstruction, at each lattice site $[i,h(i)]$ 
a particle of the type $A$ is placed with
probability $(1-P)$ and a particle of type $B$ is placed with probability
$P$. Particles $A$ are considered to be susceptible to the attack, while
particles $B$ no.
Initially all columns are filled with the same amount of atoms, so the 
the simulations start with a flat surface at $t = 0$.
The selectivity of the deconstruction 
process is simulated in the following way. 
A column $K$ is picked at random. The local curvature
$[h_{i+1}+h_{i-1}-2 h_i]$ is calculated for $i=K$, $i=K+1$ and $i=K-1$.
The particle at the top of the column with the {\em smaller} curvature is removed 
($h_K \to h_K - 1$) if it is of type $A$, otherwise it is not removed. 
However, if the local
curvature is smaller than a certain threshold $\theta$ the particle
is removed, no matter which 
type it is.~\footnote{In this way, the possibility of {\it lateral} attack
is taken into account. For instance, it could happen that a
particle $B$ may remain at the top of a long column. In a real situation,
lateral attack could detach $A$ particles underneath, leaving particle 
$B$ isolated from the interface.} 
A value of $\theta = -1$ and periodic boundary conditions were used.
It is worth to notice that for P=0, i.e., just particles of 
type $A$ in the system, the model turns to be the SC model,
the reciprocal of the LC model.

%% =====================================================================
\vspace*{1pt}
\section{Results and Discussion}
\vspace*{-0.5pt}
\noindent
Main results are summarized by the plots presented in Figs.1 and 2.
In both figures, ensemble averages where taken over 500 independent runs. 
A value of $L=2^{11}$ was used in Fig.1, while $L=2^9$ was used
in Fig.2.
As it was numerically tested, there is a critical probability
that pinn the interface at $P=P_c \cong 0.14$. A value of
$P$ slightly under $P_c$, $P=0.13$, was used in order to study 
the properties of the interface near the pinning transition. 
It is worth to notice that the critical probability has a 
different value than those corresponding to the directed percolation
transition, which is the effect responsible of the pinning transition
in interfaces invading porous media.
It can be thought that the difference is due to the fact that, 
in the model presented here, the interface becomes pinned when {\it two} 
simultaneous conditions are fulfilled: the
local curvature is greater than $\theta$ {\it and}
all the particles at the interface are of type $B$. 
On the other hand, in growth models on porous media it is enough to have
a lateral directed percolation effect in order to pinn the interface.

In Fig.1, values of the exponent $\beta$ as a function of time
are presented. The figure was constructed by calculating,
at each time $t$, the slope of the roughness
as $[log(w_{t+\Delta t}) - log(w_{t})]/ \Delta t$.
In particular, $\Delta t = 10$ was used.
This method for determining the exponent $\beta$
is known as the {\it consecutive slopes} method.~\cite{bs-b}
It can be seen that the calculated values fluctuate 
between $0.6$ and $0.7$ for the first $1000$ layers removed.
The average value of the exponent calculated up to $t=1000$ is
$\beta=0.676$. For comparison, the
corresponding exponent for the advance of an interface in 
porous media was found to be $\beta=0.7$.~\cite{bul}

In Fig.2 the a plot of the structure factor $S(k)$ is shown.
As stated above, for $t \gg L^z$ the structure factor behaves
has $S(k) \sim k^{\gamma}$ with $\gamma = 2 \alpha + 1$.
The slope of a line trough the points was found to be 
$-3.27 \pm 0.0085$, which gives a value of $\alpha \cong 1.135$. 
In this case the value of the exponent is quite different from 
the corresponding to interfaces advancing in porous 
media, $\alpha = 0.7$.~\cite{bul}

The values of $\alpha$ and $\beta$ indicate that the exponent $z$ is
$z = \alpha / \beta \cong 1.8$ and that the scaling relation 
$\alpha + z = 2$ is not fulfilled. Although, due to the strong 
nonlinear effects present in the evolution of the
interface, it is not expected to hold.

Final, it is pointed out that same kind of simulations were done
using $P=0.0$. In this case the
disorder in the bulk has no influence, since there are just one
type of particles in the system. Then the model is equivalent
to the SC or the RV models, which in turn are the reciprocal 
of the LC or WV models.
The equivalence is confirmed by the values of the exponents 
obtained for the no-disorder case: $\beta = 0.353$ and $\alpha = 1.385$, 
in good agreement with
the previously reported values, $\beta=0.365$ and $\alpha=1.4$, for
the WV model.~\cite{wolf} 

%% ======================================================================
\begin{figure}[htbp] 
%%ORIGINAL SIZE: width=1.4TRUEIN; height=1.5TRUEIN
\vspace*{13pt}
\centerline{\psfig{file=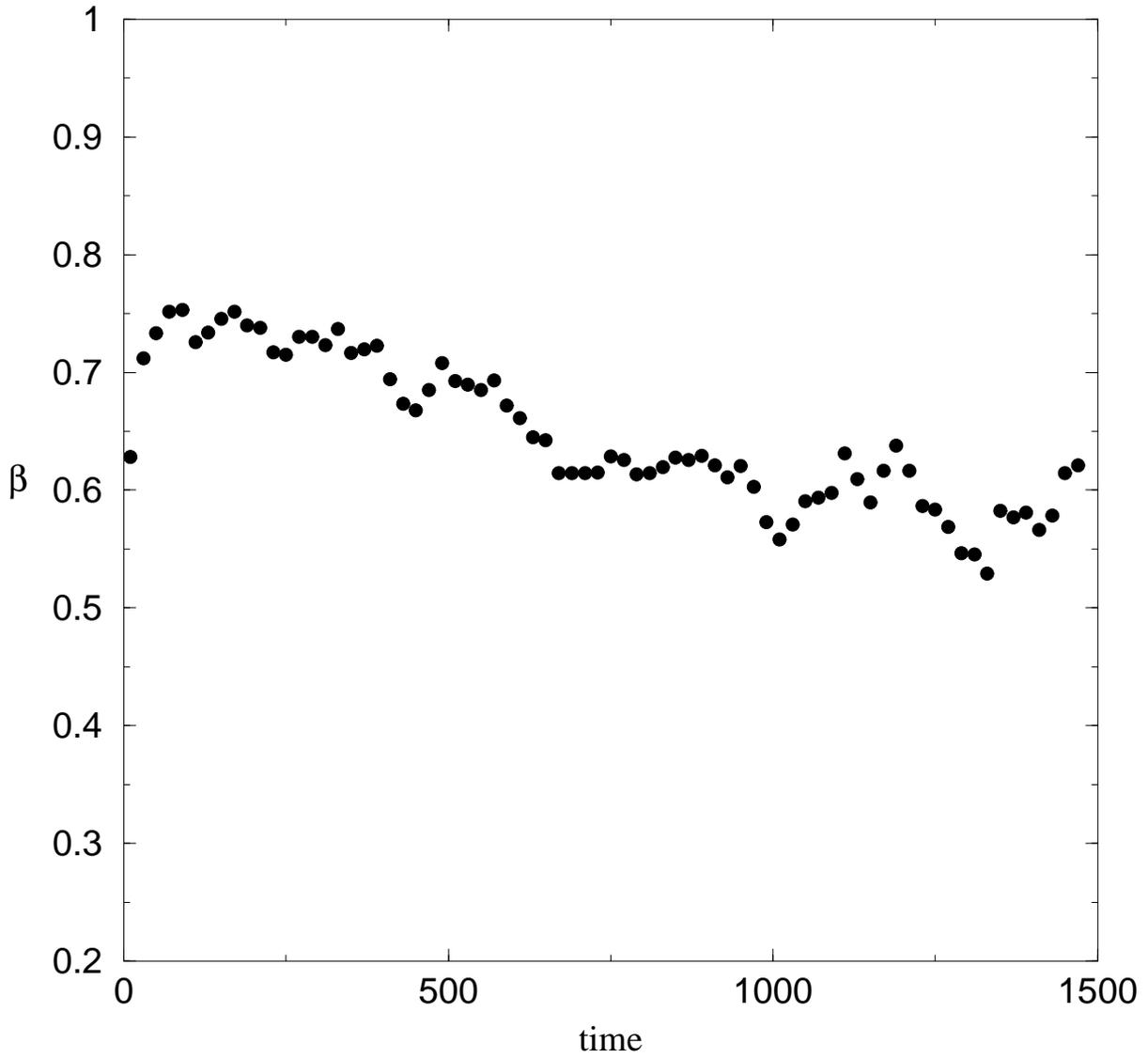}} %100 percent
\vspace*{13pt}
\caption{Values of the exponent $\beta$ as a function of
time. The average value, between $t=100$ and $t=1000$ is $0.676$.
Values of $L=2^{11}$ and $P=0.13$ were used.}
\end{figure}

\begin{figure}[htbp] 
%%ORIGINAL SIZE: width=1.4TRUEIN; height=1.5TRUEIN
\vspace*{13pt}
\centerline{\psfig{file=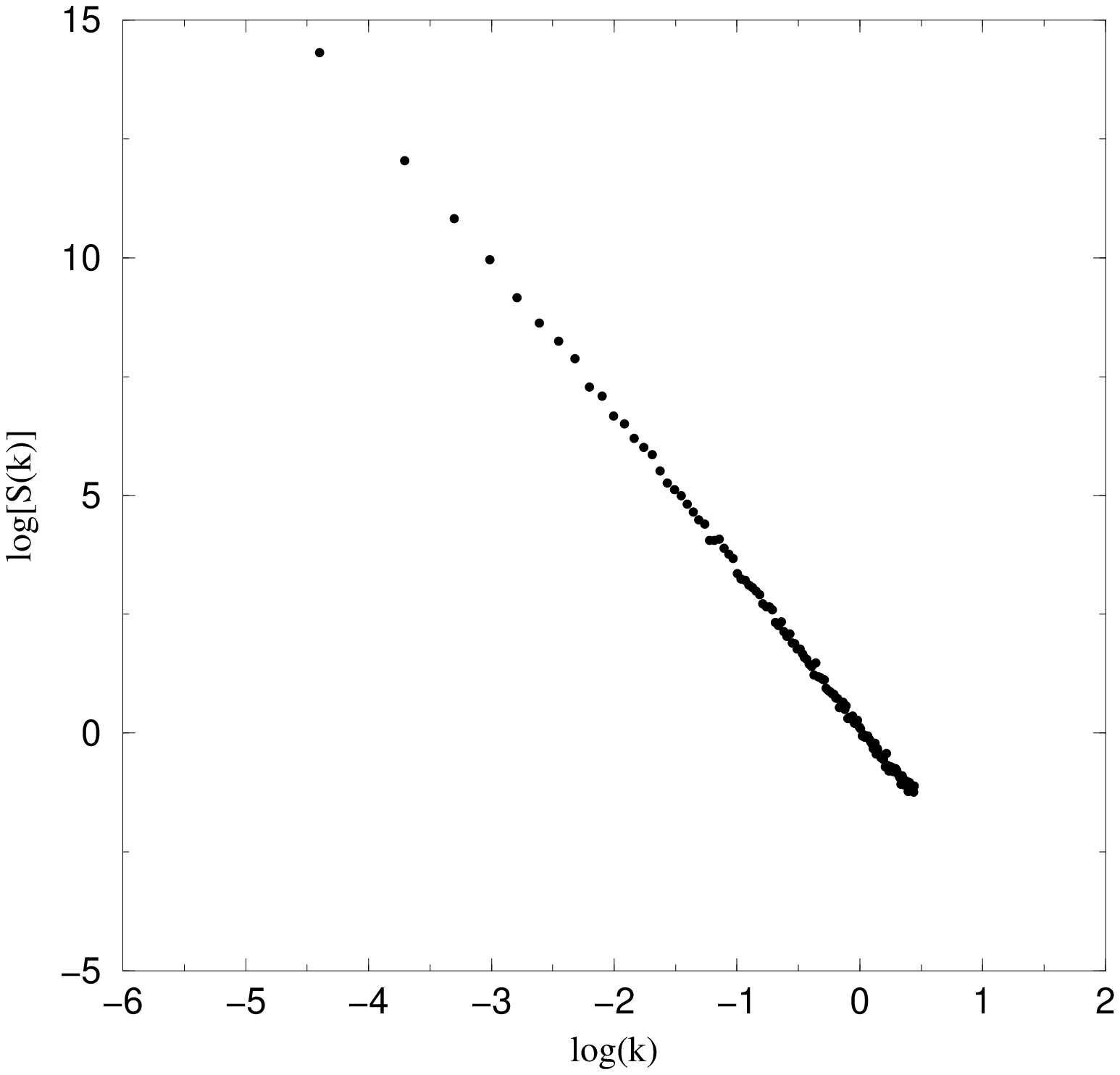}} %100 percent
\vspace*{13pt}
\caption{Plot of the structure factor $S(k)$ for $t \gg L^z$ and
$L=2^9$, with $P=0.13$. The slope of a line through the points is
$\gamma=-3.27 \pm 0.0085$.} 
\end{figure}

%% ===================================================================
\vspace*{1pt}
\section{Conclusions}
\vspace*{-0.5pt}
\noindent
A new deconstruction model has been presented. In the model
the natural heterogeneities present in all materials
of technological relevance were include. 
According to the values obtained for the scaling exponents the
model could be though to belong to a new universality class, when
the properties of the interface are studied near the
pinning transition. The values of the exponents indicate
that there is no growth model which can be associated as the
reciprocal of the model analyzed here. 
In this sense a theoretical continuum model need to be developed.
As a natural continuation of the present work, simulations on 
higher dimensions are under development.

This work was partially supported by Universidad
Nacional de Mar del Plata.

%% ===================================================================

%% =====================================================================
\end{document}